\documentclass[prd,aps,tightline,twocolumn,nofootinbib,superscriptaddress]{revtex4}
\usepackage{graphicx}
\usepackage{amssymb}
\usepackage{amsmath,bm}
\usepackage{afterpage}

\begin{document}

%\begin{titlepage}

%\begin{flushright}
%IPMU 08-0077 \\
%ICRR-Report-531
%\end{flushright}

\title{
Determination of neutrino mass hierarchy by 21 cm line and CMB B-mode 
polarization observations
}

\author{Yoshihiko Oyama}
\affiliation{The Graduate University for Advanced Studies (SOKENDAI), 1-1 Oho, Tsukuba
305-0801, Japan}

\author{Akie Shimizu}
\affiliation{The Graduate University for Advanced Studies (SOKENDAI),
1-1 Oho, Tsukuba 305-0801, Japan}

\author{Kazunori Kohri}
\affiliation{The Graduate University for Advanced Studies (SOKENDAI), 1-1 Oho, Tsukuba
305-0801, Japan}
\affiliation{Institute of Particle and Nuclear Studies, KEK, 1-1 Oho, Tsukuba
305-0801, Japan}

\date{\today}

\begin{abstract}
We focus on the ongoing and future observations for both the 21 cm
line and the CMB B-mode polarization produced by a CMB lensing, and
study their sensitivities to the effective number of neutrino species,
the total neutrino mass, and the neutrino mass hierarchy.  We find
that combining the CMB observations with future square kilometer
arrays optimized for 21 cm line such as Omniscope can determine the
%%% revised
neutrino mass hierarchy at  2$\sigma$.  We also show that a more
feasible combination of Planck + \textsc{Polarbear} and SKA can
strongly improve errors of the bounds on the total neutrino mass and
%revised version%%%%%%%%%%%%%%%%%%%%%%%%%%%%%%%%%%%%%%%%
the effective number of neutrino species to be  $ \Delta\Sigma m_{\nu}
\sim  0.12$~eV and $ \Delta N_{\nu}\sim  0.38$ at  2$\sigma$.%95 \% C.L, respectively.
%%%%%%%%%%%%%%%%%%%%%%%%%%%%%%%%%%%%%%%%
\end{abstract}

\maketitle
%\end{titlepage}

%%%%%%%%%%%%%%%%%%%%%%%%%%%%%%%%%%%%%%%%%%%%%%%%%%%%%%%%%%%%%%%%%%%%%%
\section{ Introduction }
%%%%%%%%%%%%%%%%%%%%%%%%%%%%%%%%%%%%%%%%%%%%%%%%%%%%%%%%%%%%%%%%%%%%%%

Since the discoveries of neutrino masses by Super-Kamiokande through
neutrino oscillation experiments in 1998, the standard model of
particle physics has been forced to change to theoretically include
the neutrino masses.

So far only mass-squared differences of the neutrinos have been
measured by neutrino oscillation experiments, which are reported  to
be $\Delta m^2_{21}\equiv m_{2}^2 -m_{1}^2
=7.59^{+0.19}_{-0.21}\times 10^{-5} {\rm eV}^2$~\cite{Aharmim:2008kc}
and $\Delta m^2_{32}\equiv m_{3}^2 -m_{2}^2
=2.43^{+0.13}_{-0.13}\times 10^{-3} {\rm eV}^2$~\cite{Adamson:2008zt}.
However, absolute values and their hierarchical structure (normal or
inverted) have not been obtained yet although information for them is
indispensable to build such new particle physics models.

In particle physics, some new ideas and new future experiments based
on those ideas have been proposed to observe the absolute values
and/or the hierarchy of neutrino masses, e.g., through tritium beta
decay in KATRIN experiment~\cite{KATRIN}, neutrinoless double-beta
decay~\cite{GomezCadenas:2010gs}, 
atmospheric neutrinos in the proposed iron calorimeter at INO~\cite{INO,Blennow:2012gj}
and the upgrade of the IceCube detector (PINGU) \cite{Akhmedov:2012ah},
and long-baseline oscillation experiments, e.g.,
NO$\nu$A~\cite{Ayres:2004js},
J-PARC to Korea (T2KK)~\cite{Ishitsuka:2005qi,Hagiwara:2005pe} 
and Oki island (T2KO)~\cite{Badertscher:2008bp},
and CERN to Super-Kamiokande
with high energy (5~GeV) neutrino beam~\cite{Agarwalla:2012zu}.

On the other hand, such nonzero neutrino masses affect cosmology
significantly because relativistic neutrinos prohibit the perturbation
from evolving, due to following two reasons.  First of all, in general
relativity, the density perturbation of a relativistic particle can
hardly evolve at all before it becomes nonrelativistic. Second a
relativistic neutrino erases its own density perturbation up to a
horizon scale through its free streaming at every cosmic time. By
measuring spectra of density perturbations by using observations of
cosmic microwave background (CMB) anisotropies and large-scale
structure (LSS), we could constrain the total neutrino mass
$\Sigma~m_{\nu}$ \cite{Komatsu:2008hk,Komatsu:2010fb,Reid:2009xm,
Hannestad:2010yi,Elgaroy:2003yh,
Reid:2009nq,Crotty:2004gm,Goobar:2006xz,Seljak:2004xh,Ichikawa:2004zi,
Seljak:2006bg,Fukugita:2006rm,Ichiki:2008ye,
Thomas:2009ae,RiemerSorensen:2011fe,Hamann:2010pw,Saito:2010pw}
 and the effective number of neutrino species $N_{\nu}$
\cite{Komatsu:2008hk,Komatsu:2010fb,Reid:2009xm,Reid:2009nq,Crotty:2004gm,
Pierpaoli:2003kw,Crotty:2003th,Seljak:2006bg,Hamann:2010pw}.  So far
the robust upper bound on $\Sigma~m_{\nu}$ has been obtained to be
$\Sigma m_{\nu} < 0.62$~eV (95 $\%$ C.L.)  (see
Ref.~\cite{Reid:2009xm} and references therein) by these cosmological
observations. For forecasts by future CMB observations, see also
Refs.~\cite{Lesgourgues:2005yv,dePutter:2009kn}.

In addition, by observing power spectrum of cosmological 21 cm
radiation fluctuation, we will be able to obtain independent useful
information for the neutrino
masses~\cite{Furlanetto:2006jb,Loeb:2008hg,Pritchard:2008wy,Abazajian:2011dt}. 
That is because the 21 cm radiation is emitted (1) long after the CMB epoch
(at a redshift $z \ll 10^3$ ) and (2) before an onset of the LSS
formation. The former condition (1) gives us information on
%fluctuations at larger scales, which have sensitivities on 
smaller neutrino mass ( $\lesssim 0.1 $~eV). The latter condition (2) means we
can treat only a linear regime of the matter perturbation, which can
be analytically calculated unlike the LSS case.

In actual analyses, it should be essential that we combine data of the
21 cm with that of the CMB observations because they complementary
constrain cosmological parameter spaces each other. Leaving aside
minded neutrino parameters, for example, the former is quite sensitive
to the dark energy density, but the latter is relatively insensitive
to it.  On the other hand, the former has only a mild sensitivity to
the normalization of the matter perturbation, but the latter has an
obvious sensitivity to it by definition.  In pioneering works
by~\cite{Pritchard:2008wy}, the authors tried to constrain the
neutrino mass hierarchy by combining Planck satellite with future 21
cm observations in case of relatively degenerate neutrino masses
$\Sigma m_{\nu}\sim 0.3 $~eV.

Here, however, we additionally include analyses of the CMB B-mode
polarization produced by a CMB lensing. This gives us more detailed
information on the matter power spectrum at later epochs, which means
it has better sensitivities for smaller neutrino masses down to
$\lesssim$ 0.1 eV. That is essential to distinguish the normal
hierarchy from the inverted one. In particular we adopt ongoing and
future CMB observations such as \textsc{Polarbear} and CMBPol, which
have much better sensitivities to the B-mode.  For ongoing and future
21 cm observations, we adopt MWA, SKA and Omniscope experiments. We
forecast possible allowed parameter regions for both neutrino masses
and effective number of neutrino species when we use the
above-mentioned ongoing and future observations of the 21 cm and the
CMB. In particular we propose a nice combination of neutrino masses,
$r_{\nu} = (m_3-m_1)/\Sigma m_{\nu}$ to make the mass hierarchy bring
to light as is explained in the text.

%This letter is organized as follows. In Section~\ref{sec:21cm},
%we introduce 21 cm
%line emission and how neutrino properties are constrained by the 21 cm
%observation. In Section~\ref{sec:CMB} 3 the CMB observations are explained. We show %our results in Section~\ref{sec:results}. Section~\ref{sec:conclusion} is
%devoted to our conclusion.

%%%%%%%%%%%%%%%%%FIGURE%%%%%%%%%%%%%%%%%%%%

%\begin{figure}
% \begin{center}
%   \includegraphics[width=1.0\linewidth]{fig2.ps}
%   \vspace{-1.2cm} 
%   \caption{ 
%   Allowed regions at 95$\%$ C.L. from observational light element
%   abundances in $m_{\chi}$--$\langle \sigma v \rangle$ plane.  The
%   name of each element is written in the close vicinity of the
%   line. For $^{6}$Li and $^{7}$Li,  regions sandwiched between two
%   lines are allowed, respectively. Except for lithiums, each line
%   means the upper bound. The total cross section of the annihilation
%   and its major four modes are also plotted. The calculation is
%   performed by assuming 100$\%$ $WW$ emission for simplicity. }
%   \label{fig:BBNallowed}
% \end{center}
%\end{figure}

%%%%%%%%%%%%%%%%%%%%%%%%%%%%%%%%%%%%%%%%%

%%%%%%%%%%%%%%%%%%%%%%%%%%%%%%%%%%%%%%%%%%%%%%%%%%%%%%%%%%%%%%%%%%%%%%
\section{21 cm radiation}
\label{sec:21cm}
%%%%%%%%%%%%%%%%%%%%%%%%%%%%%%%%%%%%%%%%%%%%%%%%%%%%%%%%%%%%%%%%%%%%%%
Here we briefly review basic methods to use the 21 cm line
observations as a cosmological probe.  For further details, we refer
readers to Refs.~\cite{Furlanetto:2006jb,Pritchard:2011xb}.

%%%%%%%%%%%%%%%%%%%%%%%%%%%%%%%%%%%%%%%%%%%%%%%%%%%%%%%%%%%%%%%%%%%%%%
\subsection{Power spectrum of 21 cm  radiation}
%%%%%%%%%%%%%%%%%%%%%%%%%%%%%%%%%%%%%%%%%%%%%%%%%%%%%%%%%%%%%%%%%%%%%%

The 21 cm line of the neutral hydrogen atom is emitted by  hyperfine
splitting of the 1S ground state  due to an interaction of  magnetic
moments of proton and  electron.  Spin temperature $T_{S}$ of  neutral
hydrogen gas is defined through a ratio between  number densities of
hydrogen atom in the 1S triplet and 1S singlet levels,
$n_{1}/n_{0}\equiv \left(g_{1}/g_{0} \right)\exp(-T_{\star}/T_{S})$.
where $T_{\star}\equiv hc/k_{B}\lambda_{21} = 0.068$~K with
$\lambda_{21} (\simeq 21$ cm) being the wave length of the 21 cm line
at a rest frame, and  $g_{1}/g_{0}=3$ is the ratio of  spin degeneracy
factors of the two levels.  A difference between the observed 21 cm
line brightness temperature at redshift $z$  and the CMB temperature
$T_{{\rm CMB}}$ is given by
\begin{align}
    T_{b}(\bold{x}) \approx & \ 27x_{{\rm HI}}(1+\delta_{b})
         \left( \frac{\Omega_{b}h^{2} }{0.023} \right)
         \left( \frac{0.15}{\Omega_{m}h^{2}}\frac{1+z}{10} \right)^{1/2} \nonumber \\
       & \times  \left( \frac{T_{S}-T_{{\rm CMB}}}{T_{S}} \right)
         \left( \frac{H(z)/(1+z)}{dv_{\parallel}/dr_{\parallel}} \right) \ {\rm mK},
\end{align}
where $x_{{\rm HI}}$ is the neutral fraction of hydrogen, $\delta_{b}$ is
the hydrogen density fluctuation, and $dv_{\parallel}/dr_{\parallel}$
is the gradient of the proper velocity along the line of sight due to
both the Hubble expansion and the peculiar velocity.

In general, $T_{b}$ is sensitive to details of intergalactic medium
(IGM).  However, with a few reasonable assumptions we can omit this
dependence~\cite{Madau:1996cs,Furlanetto:2006tf,Pritchard:2008da}.  At
an epoch of reionization (EOR) long after star formation begins,  X-ray
background produced by early stellar remnants has heated the IGM.
Therefore a gas kinetic temperature $T_{K}$ could be much higher than
the CMB temperature $T_{{\rm CMB}}$.  Furthermore the star formation
produces a large background of Ly$\alpha$ photons sufficient to couple
$T_{S}$ to $T_{K}$ via  the Wouthuysen-Field
effect~\cite{Wouthuysen:1952,Field:1958}.  In this scenario, we are
justified in taking $T_{{\rm CMB}} \ll T_{K} \sim T_{S} $ at  $z
\lesssim 10$, so that $T_{b}$ does not depend on $T_{S}$.

In addition, we adopt following assumptions for the EOR in the same
manner as \cite{Pritchard:2008wy,Pritchard:2009zz}.  If the IGM is
fully neutral, fluctuations of the 21 cm radiation arise only from density
fluctuations.  In this limit, we can write the power spectrum of the
21 cm line brightness fluctuation $P_{21}(\bold{k})$ as
\cite{Pritchard:2008wy}
\begin{align}
    P_{21}(\bold{k},z) =  \bar{T}_{b}^{2}(z)\left( 1+\mu^{2}\right)^{2} P_{\delta \delta}(k,z).
\end{align}
Here $P_{21}(\bold{k},z)$  is defined by  $\langle \delta
T_{b}(\bold{k})\delta T^{*}_{b}(\bold{k'}) \rangle \equiv$
$(2\pi)^{3}\delta^{3}(\bold{k}-\bold{k}')P_{21}(\bold{k})$, where
$\delta T_{b}\equiv T_{b}-\bar{T}_{b}$ is the deviation from a
spatially averaged brightness temperature $\bar{T}_{b}$, $P_{\delta
\delta}$ is the matter power spectrum, and $\mu =
\hat{\bold{k}}\cdot\hat{\bold{n}}$ is the cosine of the angle between
the wave number $\bold{k}$ and the line of sight.  In principle,
$\bar{T}_{b}$ can be calculated although it depends on the unknown
ionization and thermal history.  Therefore we treat $\bar{T}_{b}$ as a
free parameter to be measured.

The power spectrum $P_{21}({\bf k},z)$ and the comoving wave number
${\bf k}$ are not directly measured by the observations of 21 cm
radiation \cite{Mao:2008ug,Joudaki:2011sv}.  Instead, here we define
${\bf u}$ as the Fourier dual of  ${\bf \Theta} \equiv \theta_{i}
\hat{e}_{i} + \theta_{j} \hat{e}_{j} + \Delta f \hat{e}_{k}$, where
$\theta_{i}$ and $\theta_{j}$ determine an angular location on the sky
plane and $\Delta f$ shows the frequency difference from the central
redshift of a $z$ bin.  The vector ${\bf u}$ and its function
$P_{21}(\bold{u},z)$ are directly measured by the observations.
Relationships between ${\bf u}\equiv u_{i}\hat{e}_{i} + u_{j}
\hat{e}_{j} + u_{||}\hat{e}_{k}$ and ${\bf k}$ are represented by
${\bf u}_{\perp} \equiv u_{i}\hat{e}_{i} + u_{j} \hat{e}_{j} =
d_{A}(z){\bf k}_{\perp}$ $ = 2\pi {\bf L}/\lambda$, and $u_{||} = y(z)
k_{||}$.  Here "$\perp$" denotes the vector component perpendicular to
the line of sight. "$\parallel$" denotes the component in the line of
sight.  $d_{A}(z)$ is the comoving angular diameter distance to a
given redshift.  $y(z)=\lambda_{21}(1+z)^{2}/H(z)$ means the
conversion factor between comoving distance intervals and frequency
intervals $\Delta f$. ${\bf L}$ is the baseline vector of an
interferometer.  $\lambda=\lambda_{21}(1+z)$ denotes the observed wave
length of the redshifted 21 cm line.  In ${\bf u}$ space, the power
spectrum $P_{21}(\bold{u},z)$ is defined by  $\langle \delta
T_{b}(\bold{u})\delta T^{*}_{b}(\bold{u'}) \rangle  \equiv
(2\pi)^{3}\delta^{3}(\bold{u}-\bold{u}')P_{21}(\bold{u})$. Then, the
relation between $P_{21}(\bold{u},z)$ and $P_{21}(\bold{k},z)$ is
given by
\begin{align}
    P_{21}(\bold{u},z) = \frac{1}{ d^{2}_{A}(z)y(z)}P_{21}(\bold{k},z).
\end{align}
We perform our analyses in terms of $P_{21}(\bold{u},z)$ since this
quantity is directly measurable without any cosmological assumptions.
For methods of foreground removals, see also recent discussions about
independent component analysis (ICA) algorithm,
FastICA~\cite{Chapman:2012yj} which will be developed in terms of the
ongoing LOFAR observation~\cite{LOFAR}.

%%%%%%%%%%%%%%%%%%%%%%%%%%%%%%%%%%%%%%%%%%%%%%%%%%%%%%%%%%%%%%%%%%%%%%
\subsection{Effects of neutrino masses on power spectrum}
%%%%%%%%%%%%%%%%%%%%%%%%%%%%%%%%%%%%%%%%%%%%%%%%%%%%%%%%%%%%%%%%%%%%%%%
The massive neutrinos affect the growth of the matter density
perturbation mainly due to following two physical
mechanisms.~\cite{Lesgourgues:2006nd}.  First of all, a massive
neutrino $\nu_{i}$ (even with its light mass $m_{\nu_{i}} \lesssim0.3$
eV ) becomes nonrelativistic at $T \sim m_{\nu_{i}}$ and has
contributed to the energy density of cold dark matter (CDM), which
changes the matter-radiation equality epoch and has changed an
expansion rate of the universe since that time. When we fix the total
mass of neutrinos $\Sigma m_{\nu}~(\lesssim 0.3~{\rm eV})$, only the
latter effect is effective. Second, the matter density perturbations
on small scales can be suppressed due to the neutrinos'
free-streaming.  As long as neutrinos are relativistic, they travel at
speed of light, and their free-streaming scales are approximately
equal to the Hubble horizon. Then the free-streaming effect erases
their own perturbations within such scales.

Compared with the standard $\Lambda$CDM models where three massless
active neutrinos are assumed, we will consider two more freedoms.
First one is an introduction of the effective number of neutrino
species $N_{\nu}$, which counts generations of relativistic neutrinos
before the matter-radiation equality epoch and should not be equal to
three. Second one is the neutrino mass hierarchy. It is clear that a
change of $N_{\nu}$ affects the epoch of matter-radiation equality.
On the other hand, the neutrino mass hierarchy affects both the
free-streaming scales and the expansion rate as was mentioned
above~\cite{Lesgourgues:2004ps}.  In terms of the observations of the
21 cm signal, the minimum cutoff of the wave number is given by
$k_{{\rm min}}=2\pi/(yB)\sim 6\times10^{-2} h{\rm Mpc}^{-1}$ (see Sec
\ref{subsec:Forecast}) while the wave number corresponding to the
neutrino free-streaming scale is $k_{{\rm free}}\lesssim 10^{-2} h{\rm
Mpc}^{-1}$.  Therefore the main feature of the modification of the matter
density fluctuation due to the change of the mass hierarchy comes from
the modification of the cosmic expansion when we fix the total matter
density at the present time.

%%%%%%%%%%%%%%%%%%%%%%%%%%%%%%%%%%%%%%%%%%%%%%%%%%%%%%%%%%%%%%%%%%%%%%%%%%%%%%%%%%%%%%%
\subsection{Forecasting methods and interferometers}\label{subsec:Forecast}
%%%%%%%%%%%%%%%%%%%%%%%%%%%%%%%%%%%%%%%%%%%%%%%%%%%%%%%%%%%%%%%%%%%%%%%%%%%%%%%%%%%%%%%
Here we summarize future observations of the 21 cm signals emitted at
the EOR.  We also provide a brief review of the Fisher matrix
formalism for the 21 cm observations. We consider MWA \cite{MWA}, SKA
\cite{SKA} and Omniscope \cite{Tegmark:2009kv} for future
observations. The summary of the detailed specifications is listed in
Table \ref{table:interferometers}.  Each interferometer has its own
different noise power spectrum,
% $P_{N}(\bold{u}_{\perp},z)$, 
%%
\begin{align}
    P_{N}(\bold{u}_{\perp},z) = \left( \frac{\lambda^{2}(z)T_{{\rm sys}}(z)}{A_{e}(z)} \right)^{2} 
                                \frac{1}{t_{0}n(u_{\perp})},
\end{align}
which affects  sensitivities to the 21 cm signals. Here $T_{{\rm sys}}
\simeq 280[(1+z)/7.4]^{2.3}$K is the system
temperature~\cite{Wyithe:2007if}, $t_{0}$ is the total observation
time, and $A_{e}$ is the effective collecting area of each antenna
tile.  The effect of the configuration of the antennae is encoded in
the number density of baseline $n(u_{\perp})$.   In order to calculate
$n(u_{\perp})$, we have to assume  a realization of antenna density
profiles for each interferometer.  For MWA, we take 500 antennae
distributed with a filled nucleus of radius 20 m surrounded by the
remainder of the antennae distributed with an $r^{-2}$ antenna density
profile out to 750 m \cite{Bowman:2005cr}.  For SKA, we distribute
20\% of a total of 5000 antennae within a 1 km radius  and take the
antennae distributed with a nucleus surrounded by an $r^{-2}$ antenna
density profile in the same way as those of MWA.  These antennae are
surrounded  by a further 30\% of the total antennae in a uniform
density annulus of outer radius 6 km \cite{Pritchard:2008wy}. The
remainder of the antennae is distributed at  larger distances
sparsely to be useful for power spectrum measurements.  Finally, we
consider Omniscope that is a future square-kilometer collecting area
array optimized for observations of the 21 cm signal.  In case of
Omniscope, we take all of antennae distributed with a filled nucleus
according to \cite{Mao:2008ug}.

To forecast 1 $\sigma$ errors of cosmological parameters, we use the
Fisher matrix formalism \cite{Tegmark:1996bz}.  For the observations
of the 21 cm signal,  the Fisher matrix for cosmological parameters
$p_{i}$ is expressed as \cite{McQuinn:2005hk}
\begin{align}
   \bm{ \mathrm{F}}_{ij}^{21{\rm cm} }
                   = \sum_{{\rm pixels}} \frac{1}{[\delta P_{21}(\bold{u})]^{2}}
                     \left( \frac{\partial P_{21}(\bold{u})}{\partial p_{i}} \right)
                     \left( \frac{\partial P_{21}(\bold{u})}{\partial p_{j}} \right),
\end{align}
where we sum only  over half the Fourier space.  The Fisher matrix
determines the errors of the parameter $p_{i}$ to be
\begin{align}
    \Delta p_{i} \geq \sqrt{(\bm{\mathrm{F}}^{-1})_{ii}}.
\end{align}
The error of the power spectrum measurement  $\delta P_{21}({\bf u})$
in a pixel at ${\bf u}$ consists of a sum of the sample variance and
the thermal detector noise.  It is expressed as
\begin{align}
    \delta P_{21}({\bf u}) = \frac{P_{21}({\bf u})+ P_{N}(u_{\perp})}{N_{c}^{1/2}},
\end{align}
where $N_{c}=2\pi k_{\perp} \Delta k_{\perp}\Delta k_{\parallel} V(z)
/(2\pi)^{3}$ is the number of independent modes in an annulus summing
over the azimuthal angle, $V(z)=d_{A}(z)^{2}y(z)B\times $FOV is the
survey volume, $B$ is the bandwidth, and  FOV ($\approx
\lambda^{2}/A_{e}$) denotes the field of view of the interferometer.
For each experiment, we take account of the presence of foregrounds
and adopt a cutoff at $2\pi/(yB) \leq k_{\parallel}$
\cite{McQuinn:2005hk}.  We also take a maximum value of $k$ to be
$k_{{\rm max}}=3 h{\rm Mpc}^{-1}$  beyond which  nonlinear effects
become important  and exclude all information for $k_{{\rm max}} < k$.

For each experiment, we assume a specific redshift range as
follows~\cite{Pritchard:2009zz}.  We consider  Ly-$\alpha$ forests in
absorption spectra of quasars and assume that reionization occurred
sharply at $z=7.5$.  For an upper limit on the accessible redshift
range, we take it to be $z \lesssim 10$ because of increasing
foregrounds and uncertainty in the spin temperature at higher
redshifts.  For the above reasons, we assume that the observed
redshift range of EOR is $7.8-10.2$.  Only for  MWA, we assume a
single redshift slice centered at $z=8$.

%%%%%%%%%%%%%%%%%%%Table of Interferometers%%%%%%%%%%%%%%%%%%%%%%%%%%%%%%%%%%%%%%%%%%%%%%%%%%

\begin{table}[t]
\begin{center}  
\begin{tabular}{ccccccc} \hline \hline
   
  Experiment  &  $N_{ant}$ & $A_{e}(z=8)$   & $L_{{\rm min}}$ & $L_{{\rm max}}$ & FOV   & $z$ \\
              &            & [$m^{2}$] & [m]       & [km]      & [deg$^{2}$]       &   \\ \hline \hline
  MWA         & 500        & 14        & 4         & 1.5       &$\pi 16^{2}$       & 7.8-8.2\\
  SKA         & 5000       & 120       & 10        &  5        &$\pi 5.6^{2}$      & 7.8-10.2\\ 
  Omniscope   & $10^{6}$   & 1         & 1         &  1        &$2.1\times 10^{4}$ & 7.8-10.2\\
  \hline \hline 
  \end{tabular}
  \caption{Specifications for each interferometers.
           $L_{{\rm min}}$ ($L_{{\rm max}}$) is the minimum (maximum) baseline.
           For MWA, we assume a single redshift slice centered at $z=8$.
           For SKA and Omniscope, the observed redshift range is $z=7.8-10.3$, and
           we divide the range into five redshift slices with thickness $\Delta z \approx 0.5$.
           For each experiment, bandwidth is $B=8$ MHz, 
           %and observation time is 8000 hour.
           and we assumed observations for 8000 h on two places in the sky.
           We assume that the effective collecting area $A_{e}$ is 
           proportional to $\lambda^{2}$ for MWA and SKA.
           For Omniscope, both $A_{e}$ and FOV are fixed.
           \label{table:interferometers}}
\end{center}
\end{table}

%%%%%%%%%%%%%%%%%%%%%%%%%%%%%%%%%%%%%%%%%%%%%%%%%%%%%%%%%%%%%%%%%%%%%5

When we calculate the Fisher matrix, we choose the following basic set
of cosmological parameters: the energy density of  matter
$\Omega_{m}h^{2}$, baryon  $\Omega_{b}h^{2}$, dark energy
$\Omega_{\Lambda}$,  the scalar spectral index $n_{s}$, the scalar
fluctuation amplitude $A_{s}$ (the pivot scale is taken to be $k_{{\rm
pivot}}=$ $0.002 \ {\rm Mpc}^{-1}$), the reionization optical depth
$\tau$, Helium fraction $Y_{{\rm He}}$, and the total neutrino mass
$\Sigma m_{\nu} = m_{1}+m_{2}+m_{3}$. Fiducial values of these
parameters (except for $\Sigma m_{\nu}$) are adopted to be
$(\Omega_{m}h^{2},\Omega_{b}h^{2},\Omega_{\Lambda},n_{s},A_{s},\tau,Y_{\rm
He}) = (0.147,0.023,0.7, 0.95,24\times 10^{-10},0.1,0.24)$.
We set a
range of the fiducial value of $\Sigma m_{\nu}$ to be $\Sigma m_{\nu}
= 0.05 - 0.3 \ {\rm eV}$.  Besides these parameters, the brightness
temperature of 21 cm radiation $\bar{T}_{b}(z)$ can be taken as a free
parameter.  In this study,  we adopt the fiducial values of
$\bar{T}_{b}(z)$ to be $\left( \bar{T}_{b}(8), \bar{T}_{b}(8.5),
  \bar{T}_{b}(9), \bar{T}_{b}(9.5), \bar{T}_{b}(10)\right)
=\left(26,26,27,27,28 \right)$ in units of mK.

Additionally, we separately study  following two cases:
\begin{description}
  \item[(A) Effective number of neutrino species]
\end{description}

We add one more parameter of the effective number of neutrino species
$N_{\nu}$ to the fiducial set of the parameters.  The fiducial value
of this parameter is set to be $N_{\nu}=3.04$.
%
%%%%revised version
In this analysis, we assumed three species of massive neutrinos + 
an extra relativistic component.
%%%%%%%%%%%%%%%%%%%%%%%%%%%%

\begin{description}
\item[(B) Neutrino mass hierarchy]
\end{description}

In a cosmological context, many different parameterizations of the
mass hierarchy have been proposed
\cite{Takada:2005si,Slosar:2006xb,DeBernardis:2009di,Jimenez:2010ev}.
We adopt $r_{\nu}\equiv (m_{3} - m_{1})/\Sigma
m_{\nu}$~\cite{Jimenez:2010ev} as an additional parameter to nicely
discriminate the true neutrino-mass hierarchy pattern from the other
between the normal and inverted hierarchies. The normal and inverted
mass hierarchies mean $m_{1}<m_{2} \ll m_{3}$ and $m_{3} \ll
m_{1}<m_{2}$, respectively. We add $r_{\nu}$ to the fiducial set of
the parameters.  $r_{\nu}$ becomes positive (negative) for the normal
(inverted) hierarchy. It should be a remarkably nice point that the
difference between $r_{\nu}$'s of these two hierarchies becomes larger
as the total mass $\Sigma m_{\nu}$ becomes smaller. Therefore
$r_{\nu}$ is particularly useful for distinguishing the mass
hierarchy.  In Fig \ref{fig:hie} we plot behaviors of $r_{\nu}$ as a
function of $\Sigma m_{\nu}$. Note that there is a lowest value of
$\Sigma m_{\nu}$ which depends on a type of the hierarchies by the
neutrino oscillation experiments, i.e., $\sim$0.1 eV for the inverted
hierarchy and $\sim$0.05 eV for the normal hierarchy. Therefore, if we
could obtain a clear constraint like 0.05~eV~$ \le \Sigma m_{\nu} \ll
$~0.10~eV, the hierarchy should be obviously normal without any
ambiguities. As will be shown later, however, we can discriminate the
hierarchy even when 0.10~eV~$ \lesssim \Sigma m_{\nu}$.

%%%%%%%%%%%%%%%%%%%%%%%%%%%%%%%%%%%%%%%%%%%%%%%%%%%%%%%%%%5

%%%%%%%%%%%%%%%%%%%%%%%%%%%%%%%%%%%%%%%%%%%%%%%%%%%%%%%%%%%%%%%%%%%%%
\section{CMB }
\label{sec:CMB}
%%%%%%%%%%%%%%%%%%%%%%%%%%%%%%%%%%%%%%%%%%%%%%%%%%%%%%%%%%%%%%%%%%%%%%

%%%%%%%%%%%%%%%%%%%%%%%%%%%%%%%%%%%%%%%%%%%%%%%%%%%%%%%%%%%%%%%%%%%%%
\subsection{CMB and neutrino}
%%%%%%%%%%%%%%%%%%%%%%%%%%%%%%%%%%%%%%%%%%%%%%%%%%%%%%%%%%%%%%%%%%%%%

CMB power spectra are sensitive to neutrino masses.   There are three
effects that provide detectable signals for the neutrino masses: 
(1) the transition from relativistic neutrino to nonrelativistic one, (2)
smoothing of the matter perturbation by its free-streaming in small
scales, and (3) variation of lensed CMB power spectra.  Future CMB
experiments are expected to set stringent constraints on the sum of
neutrino masses~\cite{Lesgourgues:2006nd,Wong:2011ip}.  In particular,
the last effect is unique in the CMB B-mode polarization produced by a
CMB lensing. Here  we propose to combine the CMB experiments with the
21 cm line observations.  As we will see in Section~\ref{sec:results}, the
combined approach resolves degeneracy among some key cosmological
parameters and is more powerful than individual CMB measurements.  In
addition, it is notable that we are able to detect the effective
number of neutrino species~\cite{Chacko:2003dt} and determine the
neutrino mass hierarchy.

%%%%%%%%%%%%%%%%%%%%%%%%%%%%%%%%%%%%%%%%%%%%%%%%%%%%%%%%%%%%%%%%%%%%%
\subsection{Sensitivity and Analysis of the CMB experiments}
%%%%%%%%%%%%%%%%%%%%%%%%%%%%%%%%%%%%%%%%%%%%%%%%%%%%%%%%%%%%%%%%%%%%%

In this study, we choose Planck~\cite{Lesgourgues:2005yv},
\textsc{Polarbear}~\cite{paper:POLARBEAR} and
CMBPol~\cite{Baumann:2008aq} as examples of CMB experiments.
Experimental specifications we assumed are summarized in
Table~\ref{tab:sensitivity}.

%%%%%%%%%%%%%%%%%%%%%%%%%%%%%%%%%%%%%%%%%%%%%%%%%%%%%%%%%%%%%%%%%%%%%%
\begin{table}[t]
\begin{center}
\begin{tabular}{cccccc}
\hline
\hline
\shortstack{Experiment\\ \,}&
\shortstack{$\nu$\\ $[\mathrm{GHz}]$}&
\shortstack{$\Delta _{\mathrm{TT}}$\\ $[\mathrm{\mu K-'}]$}&
\shortstack{$\Delta _{\mathrm{PP}}$\\ $[\mathrm{\mu K-'}]$}& 
\shortstack{$\theta_{\mathrm{FWHM}}$\\ $[\mathrm{'}]$}&
\shortstack{$f_{\mathrm{sky}}$\\ $ $}
\\
\hline
\hline
Planck \cite{Lesgourgues:2005yv}&
70&
137&
195&
14&
0.65\\
&
100&
64.6&
104&
9.5&
0.65\\
&
143&
42.6&
80.9&
7.1&
0.65\\
\hline
\textsc{Polarbear} \cite{paper:POLARBEAR}&
150&
-&
8&
3.5&
0.017\\
\hline
CMBPol (EPIC-2m) \cite{Baumann:2008aq}&
70&
2.96&
4.19&
11&
0.65\\
&
100&
2.29&
3.24&
8&
0.65\\
&
150&
2.21&
3.13&
5&
0.65\\
\hline
\hline
\end{tabular}
\caption{Experimental specifications of Planck, \textsc{Polarbear}
and CMBPol assumed in this study. Here $\nu$ is the observation
frequency, $\Delta_{\rm TT}$ is the temperature sensitivity per $1'\times1'$
pixel, $\Delta_{\rm PP}$ is the polarization sensitivity per $1'\times1'$ pixel,
$\theta_{\rm FWHM}$ is the angular resolution defined as the full width at
half-maximum, and $f_{\rm sky}$ is the observed fraction of the sky.  We
use $\ell_{\rm max} = 2000$ for \textsc{Polarbear}, and $\ell_{\rm max} = 2500$ for Planck
and CMBPol.}
\label{tab:sensitivity}
\end{center}
\end{table}
%%%%%%%%%%%%%%%%%%%%%%%%%%%%%%%%%%%%%%%%%%%%%%%%%%%%%%%%%%%%%%%%%%%%%

In our analysis for the CMB, we also take the same fiducial model
$(\Omega_{m}h^2,\Omega_{}h^2,\Omega_{\Lambda},
n_s, A_s, \tau, Y_{\mathrm{He}})$ as that of the 21 cm line experiments
(see previous section). We evaluate errors of cosmological parameters
by using Fisher matrix, which  is given  by~\cite{Tegmark:1996bz}
%%%%%%%%%%%%%%%%%%%%%%%%%%%%%%%%%%%%%%%%%%%%%%%%%%%%%%%%%%%%%%%%%%%%%%
\begin{align}
\bm{\mathrm{F}}_{ij}^{\rm CMB} = &\sum_{l}
\frac{\left( 2\ell+1\right)}{2}
f_{\mathrm{sky}} \notag \\
&\times
\mathrm{Trace}
\left[
  \bm{\mathrm{C}}_{\ell}^{-1}
  \frac{\partial\bm{\mathrm{C}}_{\ell}}{\partial p_i}
  \bm{\mathrm{C}}_{\ell}^{-1}
  \frac{\partial \bm{\mathrm{C}}_{\ell}}{\partial p_j}
\right]. \label{eq:Fisher}
\end{align}

%%%%%%%%%%%%%%%%%%%%%%%%%%%%%%%%%%%%%%%%%%%%%%%%%%%%%%%%%%%%%%%%%%%%%%
Here $\bm{\mathrm{C}}_{\ell}$ is a covariance matrix constructed by
using   CMB power spectra $C_{\ell}^{\mathrm{X}} \left(
  \mathrm{X}=\mathrm{TT, EE, TE} \right)$, deflection angle spectrum
$C_{\ell}^{\mathrm{dd}}$, cross correlation between temperature and
deflection angle $C_{\ell}^{\mathrm{Td}}$, and noise power spectra
$N_{\ell}^{\mathrm{X}}$ and $N_{\ell}^{\mathrm{dd}}$, where
$C_{\ell}^{\mathrm{dd}}$ is calculated by a lensing
potential~\cite{Okamoto:2003zw} and is related with
$C_{\ell}^{\mathrm{BB}}$~\footnote{By performing a public code
HALOFIT~\cite{Lewis:1999bs,CAMB}, we have checked that modifications by including
nonlinear  effects for evolutions of the matter power spectrum are
much smaller than typical errors  in our analyses and negligible for
parameter fittings.}. 
We compute
$N_{\ell}^{\mathrm{dd}}$ by using a public code
FUTURCMB~\cite{paper:FUTURCMB} which adopts the quadratic
estimator~\cite{Okamoto:2003zw}. In this algorithm,
$N_{\ell}^{\mathrm{dd}}$ is reconstructed by $N_{\ell}^{\mathrm{Y}}$
$\left(\mathrm{Y}=\mathrm{TT,EE,BB}\right)$. The covariance matrix in
the Fisher matrix is expressed as
%%%%%%%%%%%%%%%%%%%%%%%%%%%%%%%%%%%%%%%%%%%%%%%%%%%%%%%%%%%%%%%%%%%%%%
\begin{align}
    \bm{\mathrm{C}}_{\ell} =  \left(
\begin{array}{ccc}
\! \!
C_{\ell}^{\mathrm{TT}} + N_{\ell}^{\mathrm{TT}} \! \! \! \! \! \! &
C_{\ell}^{\mathrm{TE}} \! \! \! \! \! \! &
C_{\ell}^{\mathrm{Td}} \! \! \\
\! \!  C_{\ell}^{\mathrm{TE}} \! \! \! \! \! \!  &
C_{\ell}^{\mathrm{EE}}+N_{\ell}^{\mathrm{EE}} \! \! \! \! \! \! &
0 \! \! \\
\! \!  C_{\ell}^{\mathrm{Td}}\! \! \! \! \! \! &
0 \! \! \! \! \! \! &
C_{\ell}^{\mathrm{dd}}+N_{\ell}^{\mathrm{dd}} \! \!
\end{array}
\right).
\end{align}
where $N_{\ell}^{\mathrm{Y}}$ is expressed by using both a beam size
$\sigma_{\mathrm{beam}}(\nu)=$ $\theta
_{\mathrm{FWHM}}(\nu)/\sqrt{8\ln 2}$ and an instrumental sensitivity
$\Delta _{\mathrm{Y}}(\nu)$ to be
\begin{eqnarray}
  N_{\ell}^{\mathrm{Y}}
  = \left[ 
    \sum_{\nu} \frac1{N_{\ell}^{\mathrm{Y}}(\nu)}
    \right]^{-1},
\end{eqnarray}
where
\begin{align}
N_{\ell}^{\mathrm{Y}}(\nu)=
\Delta ^2_{\mathrm{Y}}(\nu)\exp\left[\ell(\ell+1) \sigma ^2_{\mathrm{beam}}(\nu)\right].
\end{align}
For $\Delta_{\mathrm{EE}}(\nu)$ and $\Delta_{\mathrm{BB}}(\nu)$, we commonly use
$\Delta_{\mathrm{PP}}(\nu)$ listed in Table~\ref{tab:sensitivity}.
$N_{\ell}^{\mathrm{dd}}$ is calculated by
$N_{\ell}^{\mathrm{TT}}$, $N_{\ell}^{\mathrm{EE}}$, and
$N_{\ell}^{\mathrm{BB}}$.

In case of Planck and \textsc{Polarbear}, 
we combine both the experiments, 
and assume that the 1.7\% region of the whole sky is observed by both the experiments, 
and the remaining  $63.3\%(=65\%-1.7\%)$ region is observed by Planck only. 
%
%And the 1.7\% region and 63.3\% one are completely independent. 
%
%Under this assumption, 
Therefore we evaluate a total Fisher matrix $\bm{\mathrm{F}}^{\mathrm{CMB}}$ 
by summing the two Fisher matrices,
\begin{align}
\bm{\mathrm{F}}^{\mathrm{CMB}}&=\bm{\mathrm{F}}^{\mathrm{Planck}}(f_{\mathrm{sky}}=0.633)\notag \\
&+\bm{\mathrm{F}}^{\mathrm{Planck+PB}}(f_{\mathrm{sky}}=0.017),
\end{align}
where $\mathrm{\bm{F}^{Planck}}$ is the Fisher matrix of the region observed by Planck only
%($f_{\mathrm{sky}}=0.633$), 
and $\bm{\mathrm{F}}^{\mathrm{Planck+PB}}$
is that by both Planck and \textsc{Polarbear}. %($f_{\mathrm{sky}}=0.017$). 

In addition, we calculate noise power spectra 
$N_{\ell}^{\mathrm{Y,Planck+PB}}$ of the CMB polarization (Y= EE or BB)
in $\mathrm{F}^{\mathrm{Planck+PB}}$ with the following operation. 

\begin{description}
\item[(1)] $2\leq \ell < 25$, $2000<\ell \leq 2500$
\begin{align}
N_{\ell}^{\mathrm{Y,Planck + \textsc{PB}}} = N_{\ell}^{\mathrm{Y,Planck}}
\end{align}
\item[(2)] $25\leq \ell \leq 2000$
\begin{align}
N_{\ell}^{\mathrm{Y,Planck + \textsc{PB}}} 
= [1/N_{\ell}^{\mathrm{Y,Planck}} + 1/N_{\ell}^{\mathrm{Y,PB}}]^{-1}
\end{align}
%\item[(3)] $2000<\ell \leq 2500$
%\begin{align}
%N_{\ell}^{\mathrm{Y}} = N_{\ell}^{\mathrm{Planck}}
%\end{align}
\end{description}
Since \textsc{Polarbear} observes only CMB polarizations,
the temperature noise power spectrum $N_{\ell}^{\mathrm{TT,\,Planck+PB}}$ is equal to 
% $N_{\ell}^{\mathrm{TT,\,Planck}}$ . 
{\bf $N_{\ell}^{\mathrm{TT,\,Planck}}$.}

In order to combine the CMB experiments with the 21 cm line
experiments, we calculate the combined fisher matrix to be
\begin{align}
\bm{\mathrm{F}}^{\mathrm{21cm+CMB}} \simeq \bm{\mathrm{F}}^{\mathrm{\rm CMB}} + \bm{\mathrm{F}}^{\mathrm{21cm}}.
\end{align}
Here we did not use information for a possible correlation between
fluctuations of the 21 cm and the CMB.

%where we have approximated that the cosmic variance is much smaller
%than the noize at large $\ell$s where neutrino mass is important,
%which validates the equation to be just a simple summation.

%%%%%%%%%%%%%%%%%%%%%%%%%%%%%%%%%%%%%%%%%%%%%%%%%%%%%%%%%%%%%%%%%%%%%%%%%
\section{Results}
\label{sec:results}
%%%%%%%%%%%%%%%%%%%%%%%%%%%%%%%%%%%%%%%%%%%%%%%%%%%%%%%%%%%%%%%%%%%%%%%%%

%%%%%%%%%%%%%%%%%FIGURE%%%%%%%%%%%%%%%%%%%%
\begin{figure*}[t]
 \begin{center}
   \includegraphics[width=0.75\linewidth]{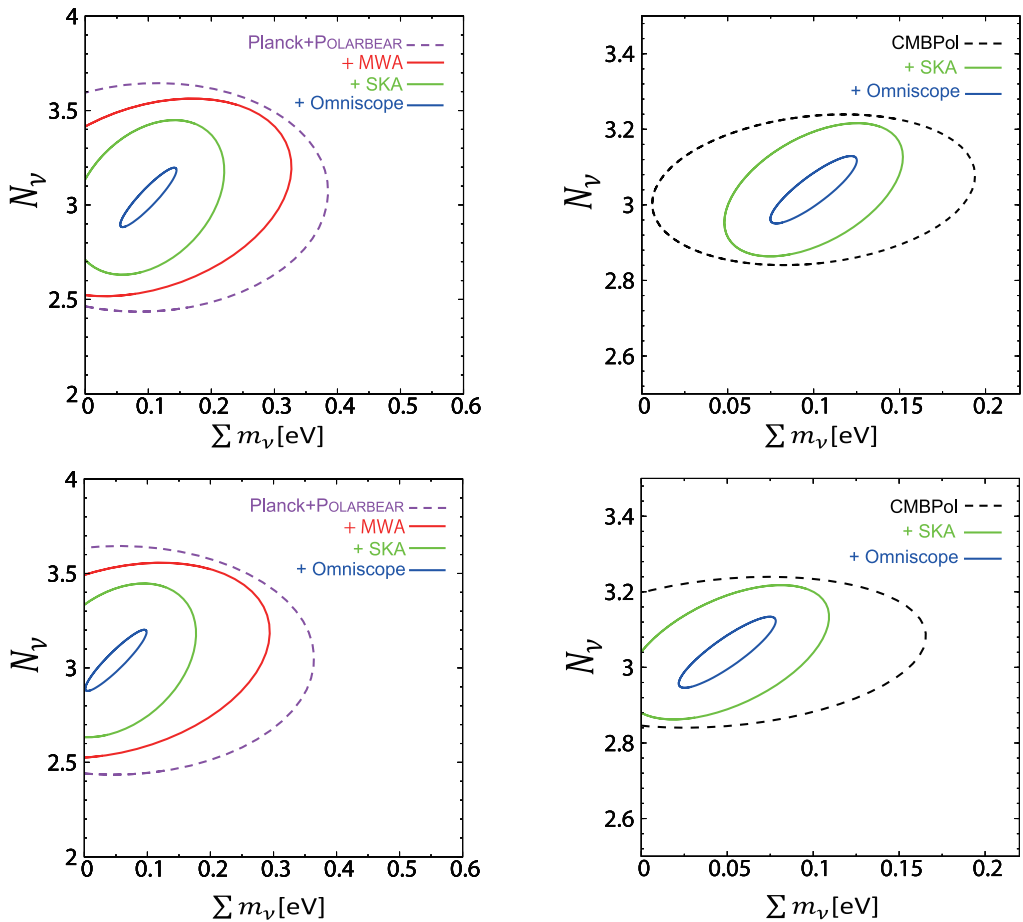}
   %\epsfile{Nnu.eps ,width=1.0\hsize}
   \caption{ 
   Contours of 
   %%%%%%revised
    90\% C.L. 
   %%%%
   forecasts in $\Sigma
   m_{\nu}$-$N_{\nu}$ plane, by adopting Planck + \textsc{Polarbear} +
   each 21 cm experiment (left two panels), or CMBPol  + each 21 cm
   experiment (right two panels). Fiducial values of neutrino
   parameters, $N_{\nu}$ and $\Sigma m_{\nu}$, are taken to be
   $N_{\nu} = 3.04$ and $\Sigma m_{\nu} = 0.1$~eV (for upper two
   panels) or $\Sigma m_{\nu} = 0.05$~   eV (for lower two panels).
   The  dashed line means the constraint obtained by only a CMB
   observation such as  Planck + \textsc{Polarbear} alone (left two
   panels), or CMBPol  alone (right two panels). The severer
   constrains are obtained by combining the CMB with a 21 cm
   observation such as MWA (outer solid, only for left panels), SKA
   (middle solid), and Omniscope (inner solid), respectively.  }
   \label{fig:Nnu}
 \end{center}
\end{figure*}
%%%%%%%%%%%%%%%%%%%%%%%%%%%%%%%%%%%%%%%%%

%%%%%%%%%%%%%%%%%%FIGURE%%%%%%%%%%%%%%%%%%%

\begin{figure*}[t]
 \begin{center}
   \includegraphics[width=1\linewidth]{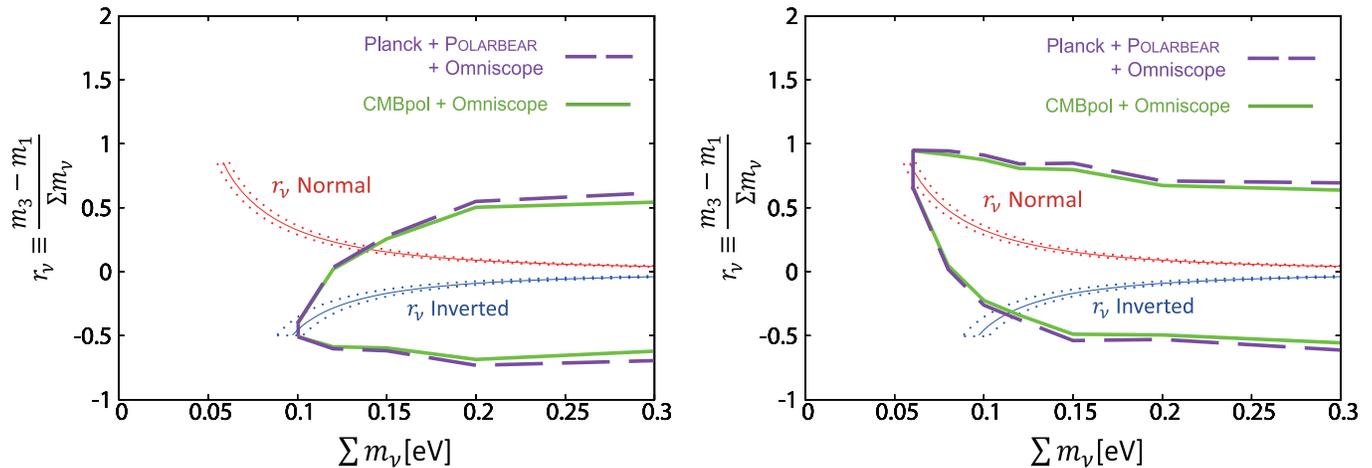} 
   \caption{
   %95\% C.L. forecasts on $r_{\nu} = (m_{3} - m_{1} )/\Sigma
   %%%%%revised 
   Forecasts of  2$\sigma$ errors on $r_{\nu} = (m_{3} - m_{1} )/\Sigma m_{\nu}$ 
   constrained by both the 21 cm and the CMB observations in
   case of the inverted hierarchy to be fiducial (left), and the
   normal hierarchy to be fiducial (right).  The constrains are
   obtained by combining Omniscope with Planck + \textsc{Polarbear}
   (thick dashed lines), and Omniscope with CMBPol (thick solid
   lines), respectively.  Allowed parameters on $r_{\nu}$ by neutrino
   oscillation experiments are plotted as two bands for the inverted
   and the normal hierarchies, respectively (the name of each
   hierarchy is written in the close vicinity of the line).  The solid
   line inside the band is the fiducial value of $r_{\nu}$ as a
   function of $\Sigma m_{\nu}$.}
   \label{fig:hie}
 \end{center}
\end{figure*}

%%%%%%%%%%%%%%%%%%%%%%%%%%%%%%%%%%%%%%%%%

%%%%%%%%%%%%revised version%%%%%%%%%%%%%
%%%%%%%%%%%%%%%%%%FIGURE%%%%%%%%%%%%%%%%%%%

\begin{figure*}[t]
 \begin{center}
   \includegraphics[width=1\linewidth]{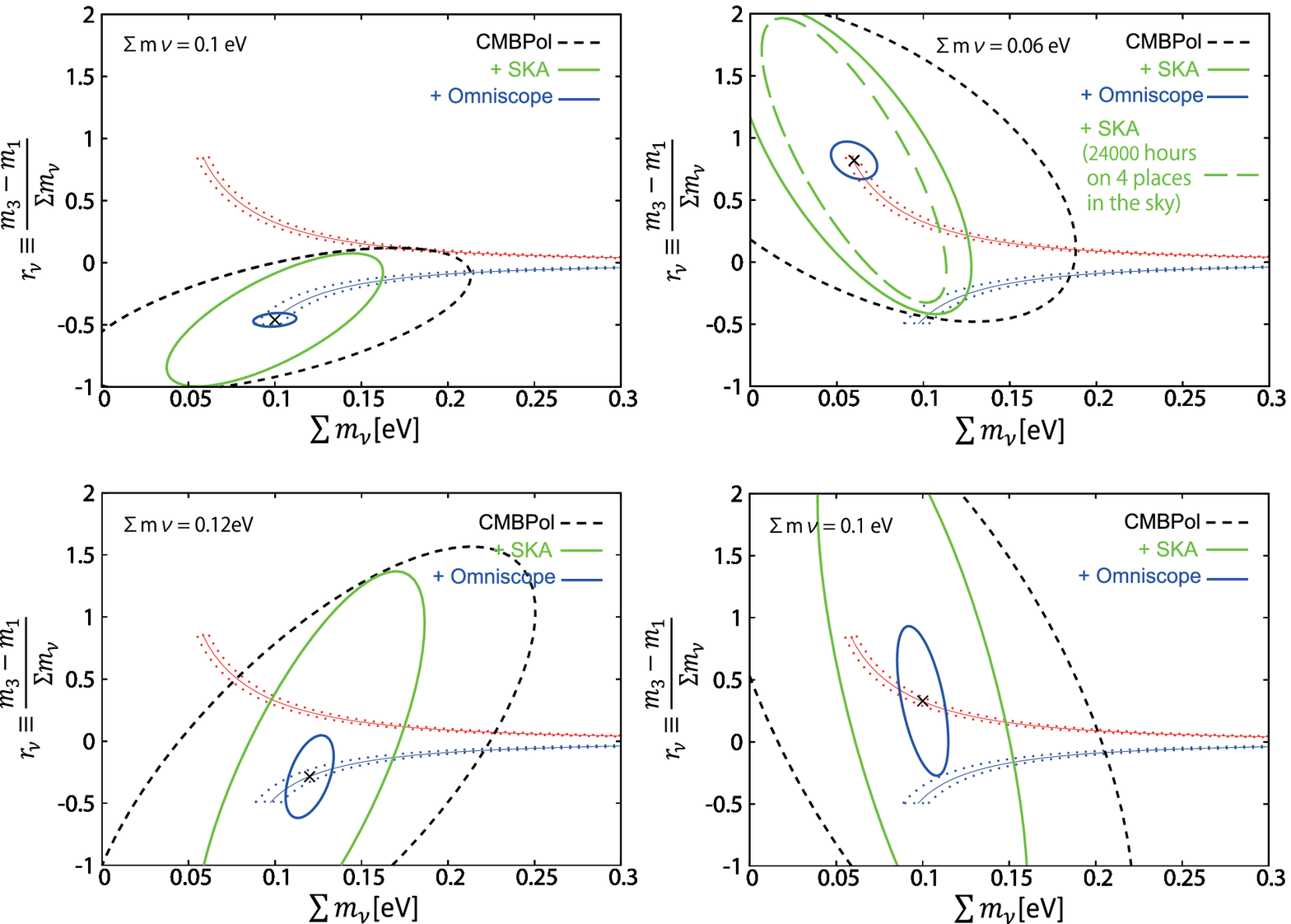} 
   \caption{
   Contours of 
   %%%%%%%%revised%
   90\% C.L.
   %%%%%%%%
   forecasts in $\Sigma
   m_{\nu}$-$r_{\nu}$ plane, by adopting CMBPol  + each 21 cm
   experiment. Fiducial value of $\Sigma m_{\nu}$ and the mass hierarchy (diagonal cross) 
   are taken to be: $\Sigma m_{\nu} = 0.1$~eV and the inverted (for left upper panel),
   $\Sigma m_{\nu} = 0.12$~eV and the inverted (for left lower panel) ,
   $\Sigma m_{\nu} = 0.06$~eV and the normal (for right upper panel),
   $\Sigma m_{\nu} = 0.1$~eV and the normal (for right lower panel).
   %The dashed line means the constraint obtained by only a CMBPol
   %%%%%revised
   The  short dashed lines  mean the constraints obtained by only a CMBPol observation,
   and the long dashed line means the one by a SKA + CMBPol observation for 24000 h on four places in the sky.
   The outer (inner) solid line means combining the CMBPol with SKA (Omniscope).
Allowed parameters on $r_{\nu}$ by neutrino
oscillation experiments are plotted 
in the same way of Fig.~\ref{fig:hie}.
}
   \label{fig:hie_ellipse}
 \end{center}
\end{figure*}

%%%%%%%%%%%%%%%%%%%%%%%%%%%%%%%%%%%%%%%%%

%%%%%%%%%%%%%%%%%%%%%%%%%%%%%%%%%%%%%%%%%%

In this section, we numerically  evaluate  how we can determine (A)
the effective number of neutrino species, and (B) the neutrino mass
hierarchy, by combining the  21 cm line observations (MWA, SKA, or
Omniscope) with the CMB experiments (Planck + \textsc{Polarbear}, or
CMBPol).  To obtain Fisher matrices we use CAMB \cite{Lewis:1999bs,CAMB} for
calculations of CMB anisotropies $C_{l}$  and matter power spectra
$P_{\delta \delta}$.

%revised version%%%%%%%%%%%%%%%%%%%%%%%%%%%%%%%%%%%%%%%%

%%%%%%%%%%%%%%%%%%%%%%%%%%%%%%%%%%%%%%%%%%%%%%%%%%%%%%%%%%%%%%%%%%%%%%
\begin{table}[t]
\begin{center}  
\begin{tabular}{lll} \hline \hline
   
                               &  $\Sigma m_{\nu}$ [eV] & $N_{\nu}$  \\
  Fiducial value               &  0.05            & 3.04       \\ \hline 
  Planck + \textsc{Polarbear}  &  0.146           & 0.282      \\
\  + MWA                       &  0.114      & 0.240       \\
\  + SKA                       &  0.0592     & 0.189      \\ 
\  + Omniscope                 &  0.0226     & 0.0753     \\
  CMBPol                       &  0.0538     & 0.0929      \\
\  + SKA                       &  0.0276     & 0.0827      \\ 
\  + Omniscope                 &  0.0131     & 0.0438       \\
  \hline \hline 
\end{tabular}
  \caption{ 1-$\sigma$ experimental uncertainties of $\Sigma m_{\nu}$
  and $N_{\nu}$, defined by  $\Delta p_{i} =
  \sqrt[]{(\bold{F}^{-1})_{ii}}$.
\label{table3}}
\end{center}
\end{table}
%revised version%%%%%%%%%%%%%%%%%%%%%%%%%%%%%%%%%%%%%%%%

%%%%%%%%%%%%%%%%%%%%%%%%%%%%%%%%%%%%%%%%%%%%%%%%%%%%%%%%%%%%%%%%%%%%%%
\subsection{Constraints on $N_{\nu}$}
%%%%%%%%%%%%%%%%%%%%%%%%%%%%%%%%%%%%%%%%%%%%%%%%%%%%%%%%%%%%%%%%%%%%%%
%
%In Fig.\ref{fig:Nnu}, we plot contours of 95\% confidence levels
%(C.L.) forecasts in $\Sigma m_{\nu}$-$N_{\nu}$ plane. 
%%%%%%%%%%revised
In Fig.\ref{fig:Nnu},  we plot contours of  90\%confidence levels
(C.L.) forecasts  in $\Sigma m_{\nu}$-$N_{\nu}$ plane. 
%%%%%%%%%%%%%%%%%%%
%
The fiducial
values of neutrino parameters, $N_{\nu}$ and $\Sigma m_{\nu}$, are
taken to be $N_{\nu} = 3.04$ and $\Sigma m_{\nu} = 0.1$~eV (upper two
panels), which corresponds to the lowest value of the inverted
hierarchy model, or $\Sigma m_{\nu} = 0.05$~eV (lower two panels),
which corresponds to the lowest value of the normal hierarchy
model. Adding the 21 cm experiments to the CMB experiment, we see that
there is a substantial improvement for the sensitivities to $\Sigma
m_{\nu}$ and $N_{\nu}$. That is because several parameter degeneracies
are broken by those combinations, e.g., in particular $T_{b}$ and $As$
were completely degenerate only in 21 cm line measurements.  Therefore
it is essential to add the CMB to the 21 cm experiment to be vital for
breaking those parameter degeneracies.

If each CMB experiment is combined with SKA or Omniscope, the
corresponding constraint can be significantly improved.  We showed
numerical values of those errors in Table~\ref{table3} in case that
the fiducial values are taken to be $N_{\nu} = 3.04$ and $\Sigma
m_{\nu} = 0.05$~eV. On the other hand, comparing those values with the
current best bounds for $\Sigma m_{\nu}$ + $N_{\nu}$ model, which give
$\Sigma m_{\nu} < 0.89$ eV and $N_{\nu} = 4.47^{+1.82}_{-1.74}$
obtained by CMB (WMAP) + HST(Hubble Space Telescope) + BAO
\cite{Hamann:2010pw}, we find that the ongoing and future 21 cm line +
the CMB observation will be able to constrain those parameters much
more severely.

The case of $\Sigma m_{\nu}=0.1$ eV to be fiducial (upper two panels)
corresponds to the lowest value for the inverted hierarchy when we use
oscillation data. Then it is notable that CMBPol + SKA can detect the
nonzero neutrino mass. Of course, Planck + \textsc{Polarbear} +
Omniscope and CMBPol + Omniscope can obviously do the same job.

On the other hand, the case of $\Sigma m_{\nu}=0.05$ eV  to be
fiducial (lower two panels), which corresponds to the lowest value for
the normal hierarchy, only Planck + \textsc{Polarbear} + Omniscope or
CMBPol + Omniscope can detect the nonzero neutrino mass.

%%%%%%%%%%%%%%%%%%%%%%%%%%%%%%%%%%%%%%%%%%%%%%%%%%%%%%%%%%%%%%%%%%%%%%
\subsection{Constraints on neutrino mass hierarchy}
%%%%%%%%%%%%%%%%%%%%%%%%%%%%%%%%%%%%%%%%%%%%%%%%%%%%%%%%%%%%%%%%%%%%%%

Next we discuss if we will be able to determine the neutrino mass
hierarchies by using those ongoing and future 21 cm and CMB
observations. 
%
%In Fig.~\ref{fig:hie} we plot 95\% C.L. on a parameter
%%%%%%%%revised
In Fig.~\ref{fig:hie} we plot  2${\bf \sigma}$ errors of the parameter 
%%%%%%%%%%%%%%
$r_{\nu}\equiv (m_{3} - m_{1} )/\Sigma m_{\nu}$ constrained by both
the 21 cm and the CMB observations in case of the inverted hierarchy
to be fiducial (left), and the normal hierarchy to be fiducial
(right). It is notable that the difference between $r_{\nu}$'s of
these two hierarchies becomes larger as the total mass $\Sigma
m_{\nu}$ becomes smaller. Therefore, $r_{\nu}$ is quite useful to
distinguish a true mass hierarchy from the other.  Allowed parameters
on $r_{\nu}$ by neutrino oscillation experiments are plotted as two
bands for the inverted and the normal hierarchies, respectively.  The
thin solid lines inside the bands are the experimental mean values by
oscillations, one of which is taken to be a corresponding fiducial
value of $r_{\nu}$ as a function of $\Sigma m_{\nu}$ in each
analysis. The constrains are obtained by combining Omniscope with
Planck + \textsc{Polarbear} (thick dashed lines) and Omniscope with
CMBPol (thick solid lines), respectively.

For $0.3 \ {\rm eV} \lesssim \Sigma m_{\nu}$ the mass eigenvalues
$m_{1}$, $m_{2}$, and $m_{3}$ are almost degenerate.  Therefore the
difference between two hierarchies has little influence on the matter
power spectrum. Therefore the constraints on $r_{\nu}$ are
significantly weak compared with the difference between them, and then
we cannot distinguish the true hierarchy from the other.

On  the  other hand  however,  the  difference  increases as  $\Sigma
m_{\nu}$ decreases down to $m_{\nu} \sim  0.1 \ {\rm eV}$. By using this
property, the  CMB  (Planck + \textsc{Polarbear} or CMBPol)  +  the 21
cm (Omniscope)  observations can  constrain the  neutrino  mass
hierarchy severely.  
%%%%%%%revised%%%%%%%%%%%%%%%%%%%%%%%%%%%%%%%%%%
%Typically errors of  $\Sigma m_{\nu}$ at around
%$\Sigma m_{\nu}=0.1$ eV is given by  $\Delta \Sigma m_{\nu}=0.012$~eV
%for Planck + \textsc{Polarbear},  and $\Delta \Sigma m_{\nu}=0.0092$~eV
%for CMBPol at 68.3\% C.L., respectively.  Therefore, the error of
%the $x$-axis is negligible compared with that of $y$-axis. 
Typically errors of  $\Sigma m_{\nu}$ at around
$\Sigma m_{\nu}=0.1$ eV are given by $ \Delta \Sigma m_{\nu}=  0.0087$~eV
for Planck + \textsc{Polarbear},  and $\Delta \Sigma m_{\nu}=0.0069$~eV
for CMBPol at  1${\bf \sigma}$, respectively.  Therefore, the error of
the $x$-axis is negligible compared with that of $y$-axis. 
%%%%%%%%%%%%%%
%
%%%revised ver%%%%%%%%%%%%%%%%%%%%%%%%%%%%%%%%%%%%%%%%
In Fig.~\ref{fig:hie_ellipse},
%
%we plot contours of 95\% C.L. in $\Sigma m_{\nu}-r_{\nu}$ plane
%%%%%revised
 we plot contours of 90\% C.L. in $\Sigma m_{\nu}-r_{\nu}$ plane 
in order to %represent these $x$-axis ($\Sigma m_{\nu}$) errors.
show errors of $\Sigma m_{\nu}$ along the $x$-axis for typical
fiducial values.
%%%%%%%%%%%%%%%%%%%%%%%%%%%%%%%%%%%%%%%%%%%%%%%%%%%%%%
%
As is
clearly shown in Fig.~\ref{fig:hie}, actually those combinations of the
observations will be able to determine   the neutrino mass hierarchy
to be inverted  or normal for $\Sigma m_{\nu}\lesssim 0.13$ eV or 
%
%revised version%%%%%%%%%%%%%%%%%%%%%%%%%%%%%%%%%%%%%%%%
%$\Sigma m_{\nu}\lesssim 0.1 \ {\rm eV}$ at 95 \% C.L.,
$\Sigma m_{\nu}\lesssim 0.1 \ {\rm eV} $ at  90 \% C.L.,
%%%%%%%%%%%%%%%%%%%%%%%%%%%%%%%%%%%%%%%
%
respectively.  Although the determination is possible only at around
$\Sigma m_{\nu}\lesssim {\cal O}(0.1) \ {\rm eV}$, those results should be
reasonable. That is  because a precise discrimination of the mass
hierarchy itself may have no meaning if the masses are highly
degenerate, i.e., if $\Sigma m_{\nu} \gg 0.1 - 0.3$~eV.

Once a clear signature $\Sigma m_{\nu} \ll 0.1 \ {\rm eV}$ were
determined by observations or experiments, it should be obvious that
the hierarchy must be normal without any ambiguities. On the other
hand if the hierarchy were inverted, we could not determine it only by
using $\Sigma m_{\nu}$.  However, it is remarkable that our method is
quite useful because we can discriminate the hierarchy from the other
even if the fiducial values were $\Sigma m_{\nu} \gtrsim 0.1$~eV for
both the normal and inverted cases.
%%
%%%%%%%%%%%%%%%%%
This is clearly shown in Fig~\ref{fig:hie_ellipse}.  In case that a
fiducial value of $\Sigma m_{\nu}$ is taken to be the lowest values in
neutrino oscillation experiments, the upper left (right) figure
indicates that even CMBPol+ SKA can discriminate the inverted (normal)
mass hierarchy from the normal (inverted) one.
%%%%%%%%%%%%%%%%%
%%

%%%%%%%%%%%%%%%%%%%%%%%%%%%%%%%%%%%%%%%%%%%%%%%%%%%%%%%%%%%%%%%%%%%%%%
\section{Conclusions}
\label{sec:conclusion}
%%%%%%%%%%%%%%%%%%%%%%%%%%%%%%%%%%%%%%%%%%%%%%%%%%%%%%%%%%%%%%%%%%%%%%

We have studied how we can constrain effective number of neutrino
species $N_{\nu}$, total neutrino masses $\Sigma m_{\nu}$, and
neutrino mass hierarchy by using the 21 cm observations (MWA, SKA, and
Omniscope) and the CMB observations (Planck, \textsc{Polarbear}, and
CMBPol). It is essential to combine the 21 cm with the CMB B-mode
polarization produced by a CMB lensing to break various degeneracies
in cosmological parameters when we perform multiple-parameter
fittings.

About the constraints on $\Sigma m_{\nu}$--$N_{\nu}$ plane, for a
fiducial value $\Sigma m_{\nu}=0.1$ eV which corresponds to the lowest
value in the inverted hierarchy,  we have found that CMBPol +
SKA, Planck + \textsc{Polarbear} + Omniscope  and  CMBPol + Omniscope can
detect the nonzero neutrino mass. For a fiducial value $\Sigma
m_{\nu}=0.05$ eV, which corresponds to the lowest value in the normal
hierarchy, Planck + \textsc{Polarbear} + Omniscope or CMBPol + Omniscope
can detect the nonzero neutrino mass.

As for the determination of the neutrino mass hierarchy, we have
proposed a new parameter $r_{\nu}= (m_{3} - m_{1} )/\Sigma m_{\nu}$
and studied how to discriminate a true hierarchy from the other by
constraining $r_{\nu}$. As was clearly shown in Fig.~\ref{fig:hie},
the combinations of the CMB (Planck + \textsc{Polarbear} or CMBPol) +
the 21 cm (Omniscope) will be able to determine the hierarchy to be
%%%%revised ver
%inverted or normal for $\Sigma m_{\nu}\lesssim 0.13$~eV or $\lesssim
%0.1$~eV at 95 \% C.L., respectively.
inverted or normal for $ \Sigma m_{\nu}\lesssim 0.13$~eV or 
$\lesssim 0.1$~eV at 2$\sigma$, respectively.
%%%%%%%%%%%%%%%%%
%
%%%%%%%%revised version%%%%%%%%%
Furthermore, 
if the fiducial value of $\Sigma m_{\nu}$ is taken to be
the lowest value in the neutrino oscillation experiments, even CMBPol +
SKA can determine the mass hierarchy.
%%%%%%%%%%%%%%%%%%%%%%%%%%%%%%%

In this study we have taken the simplified model of reionization.  In
case of more likely detailed modeling of reionization
\cite{Mao:2008ug}, it was pointed out that the constraints on
cosmological parameters may moderately change at $\sim 10 - 50$ \%.
Fortunately, this effect is comparatively small and should not be
fatal to constrain the neutrino mass hierarchy in the current
analyses.

%%%%%%%%%%%%%%%%%%%%%%%%%%%%%%%%%%%%
%\underline{\bf Acknowledgment}
%%%%%%%%%%%%%%%%%%%%%%%%%%%%%%%%%%%%
\section{Acknowledgment}

We thank M.~Hazumi, K.~Ioka, H.~Kodama, R. Nagata, T.~Namikawa, M.~Takada and
A.~Taruya for useful discussions.  This work is supported in part by
Grant-in-Aid for Scientific research from the Ministry of Education,
Science, Sports, and Culture (MEXT), Japan, No.~245516 (A.S.),
Nos.~21111006,~22244030, and ~23540327 (K.K.).

%%%%%%%%%%%%%%%%%%%%%%%%%%%%%%%%%%%%
%\section*{Appendix} \label{app}
%%%%%%%%%%%%%%%%%%%%%%%%%%%%%%%%%%%%

%\appendix

%%%%%%%%%%%%%%%%%%%%%%%%%%%%%%%%%%%%
{}
%%%%%%%%%%%%%%%%%%%%%%%%%%%%%%%%%%%%


\begin{thebibliography}{}
%%%%%%%%%%%%%%%%%%%%%%%%%%%%%%%%%%%%


\bibitem{Aharmim:2008kc} 
  B.~Aharmim {\it et al.}  [SNO Collaboration],
  %``An Independent Measurement of the Total Active B-8 Solar Neutrino Flux Using an Array of He-3 Proportional Counters at the Sudbury Neutrino Observatory,''  
Phys.\ Rev.\ Lett.\  {\bf 101}, 111301 (2008).
%  [arXiv:0806.0989 [nucl-ex]].  
%%CITATION = ARXIV:0806.0989;%%


\bibitem{Adamson:2008zt} 
  P.~Adamson {\it et al.}  [MINOS Collaboration],
  %``Measurement of Neutrino Oscillations with the MINOS Detectors in the NuMI Beam,''  
Phys.\ Rev.\ Lett.\  {\bf 101}, 131802 (2008)  [arXiv:0806.2237 [hep-ex]].  
%%CITATION = ARXIV:0806.2237;%%

\bibitem{KATRIN}
%\bibitem{Thummler:2011zz} 
  T.~Thummler [KATRIN Collaboration],
  %``Direct neutrino mass measurements,''  
Phys.\ Part.\ Nucl.\  {\bf 42}, 590 (2011);
%%CITATION = PPNUE,42,590;%%
%\bibitem{Beck:2010zzb} 
  M.~Beck [KATRIN Collaboration],
  %``The KATRIN Experiment,''  
J.\ Phys.\ Conf.\ Ser.\  {\bf 203}, 012097 (2010)  [arXiv:0910.4862
[nucl-ex]].  
%%CITATION = ARXIV:0910.4862;%%

\bibitem{GomezCadenas:2010gs} 
  J.~J.~Gomez-Cadenas, J.~Martin-Albo, M.~Sorel, P.~Ferrario, F.~Monrabal, J.~Munoz-Vidal, P.~Novella and A.~Poves,
  %``Sense and sensitivity of double beta decay experiments,''  
JCAP {\bf 1106}, 007 (2011)  [arXiv:1010.5112 [hep-ex]].  
%%CITATION = ARXIV:1010.5112;%%

\bibitem{INO} 
INO, India Based Neutrino Observatory, URL http://www.ino.tifr.res.in/ino/
%%CITATION 

%\cite{Blennow:2012gj}
\bibitem{Blennow:2012gj} 
  M.~Blennow and T.~Schwetz,
  %``Identifying the Neutrino mass Ordering with INO and NOvA,''
  arXiv:1203.3388 [hep-ph].
  %%CITATION = ARXIV:1203.3388;%%

%\cite{Akhmedov:2012ah}
\bibitem{Akhmedov:2012ah} 
  E.~K.~.Akhmedov, S.~Razzaque and A.~Y.~.Smirnov,
  %``Mass hierarchy, 2-3 mixing and CP-phase with Huge Atmospheric Neutrino Detectors,''
  arXiv:1205.7071 [hep-ph].
  %%CITATION = ARXIV:1205.7071;%%

%%%Long base line exprt%%%%%

%%%Nova%%%
%\cite{Ayres:2004js}
\bibitem{Ayres:2004js} 
  D.~S.~Ayres {\it et al.} [NOvA Collaboration],
  %``NOvA: Proposal to build a 30 kiloton off-axis detector to study nu(mu) ---> nu(e) oscillations in the NuMI beamline,''
  hep-ex/0503053.
  %%CITATION = HEP-EX/0503053;%%

%%%T2KK%%%
%\cite{Ishitsuka:2005qi}
\bibitem{Ishitsuka:2005qi} 
  M.~Ishitsuka, T.~Kajita, H.~Minakata and H.~Nunokawa,
  %``Resolving neutrino mass hierarchy and CP degeneracy by two identical detectors with different baselines,''
  Phys.\ Rev.\ D {\bf 72}, 033003 (2005)
  [hep-ph/0504026].
  %%CITATION = HEP-PH/0504026;%%

%\cite{Hagiwara:2005pe}
\bibitem{Hagiwara:2005pe} 
  K.~Hagiwara, N.~Okamura and K.~-i.~Senda,
  %``Solving the neutrino parameter degeneracy by measuring the T2K off-axis beam in Korea,''
  Phys.\ Lett.\ B {\bf 637}, 266 (2006)
  [Erratum-ibid.\ B {\bf 641}, 491 (2006)]
  [hep-ph/0504061].
  %%CITATION = HEP-PH/0504061;%%

%%%T2KO%%%
%\cite{Badertscher:2008bp}
\bibitem{Badertscher:2008bp} 
  A.~Badertscher, T.~Hasegawa, T.~Kobayashi, A.~Marchionni, A.~Meregaglia, T.~Maruyama, K.~Nishikawa and A.~Rubbia,
  %``A Possible Future Long Baseline Neutrino and Nucleon Decay Experiment with a 100 kton Liquid Argon TPC at Okinoshima using the J-PARC Neutrino Facility,''
  arXiv:0804.2111 [hep-ph].
  %%CITATION = ARXIV:0804.2111;%%


\bibitem{Agarwalla:2012zu} 
  S.~K.~Agarwalla and P.~Hernandez,
  %``Probing the Neutrino Mass Hierarchy with Super-Kamiokande,''  
arXiv:1204.4217 [hep-ph].  
%%CITATION = ARXIV:1204.4217;%%

%%%%%%%%%%%%%%%%%%%%%%%%%%%%%%


%%%%%%%%%%%%%%%%%neutrino mass bound%%%%%%%%%%%%%%%%%%%%%%%%%%%%%%%%%%%%%%%%%%%%%%%%%%%%%%%%%%%%%%

%\cite{Elgaroy:2003yh}
\bibitem{Elgaroy:2003yh} 
  O.~Elgaroy and O.~Lahav,
  %``Upper limits on neutrino masses from the 2dFGRS and WMAP: the role of priors,''  
JCAP {\bf 0304}, 004 (2003)  [astro-ph/0303089].  %%CITATION = ASTRO-PH/0303089;%%

\bibitem{Ichikawa:2004zi} 
  K.~Ichikawa, M.~Fukugita and M.~Kawasaki,
  %``Constraining neutrino masses by CMB experiments alone,''  
  Phys.\ Rev.\ D {\bf 71}, 043001 (2005).
  %  [astro-ph/0409768].  
  %%CITATION = ASTRO-PH/0409768;%%


%%%%%%%%%%% WMAP + Ly alpha + SDSSBAO %%%%

%\cite{Seljak:2004xh}
\bibitem{Seljak:2004xh} 
  U.~Seljak {\it et al.}  [SDSS Collaboration],
  %``Cosmological parameter analysis including SDSS Ly-alpha forest and galaxy bias: Constraints on the primordial spectrum of fluctuations, neutrino mass, and dark energy,''  
Phys.\ Rev.\ D {\bf 71}, 103515 (2005)  [astro-ph/0407372].  %%CITATION = ASTRO-PH/0407372;%%

%\cite{Goobar:2006xz}
\bibitem{Goobar:2006xz} 
  A.~Goobar, S.~Hannestad, E.~Mortsell and H.~Tu,
  %``A new bound on the neutrino mass from the sdss baryon acoustic peak,''  
JCAP {\bf 0606}, 019 (2006)  [astro-ph/0602155].  %%CITATION = ASTRO-PH/0602155;%%

%%%%%%%%%%%%%%%%%%%%%%%%%%%%%%%%%%%%%

\bibitem{Ichiki:2008ye} 
  K.~Ichiki, M.~Takada and T.~Takahashi,
  %``Constraints on Neutrino Masses from Weak Lensing,''  
  Phys.\ Rev.\ D {\bf 79}, 023520 (2009).
  %[arXiv:0810.4921 [astro-ph]].  
  %%CITATION = ARXIV:0810.4921;%%

%%%%%%%%%%%%%current strong bound%%%%
%\cite{Thomas:2009ae}
\bibitem{Thomas:2009ae} 
  S.~A.~Thomas, F.~B.~Abdalla and O.~Lahav,
  %``Upper Bound of 0.28eV on the Neutrino Masses from the Largest Photometric Redshift Survey,''  
Phys.\ Rev.\ Lett.\  {\bf 105}, 031301 (2010)  [arXiv:0911.5291 [astro-ph.CO]].  %%CITATION = ARXIV:0911.5291;%%

%\cite{RiemerSorensen:2011fe}
\bibitem{RiemerSorensen:2011fe} 
  S.~Riemer-Sorensen, C.~Blake, D.~Parkinson, T.~M.~Davis, S.~Brough, M.~Colless, C.~Contreras and W.~Couch {\it et al.},
  %``The WiggleZ Dark Energy Survey: Cosmological neutrino mass constraint from blue high-redshift galaxies,''  
Phys.\ Rev.\ D {\bf 85}, 081101 (2012)  [arXiv:1112.4940 [astro-ph.CO]].  %%CITATION = ARXIV:1112.4940;%%

%%%%%%%%%%%%%%%%%%%%%%%%%%%%%%%%%%%%%

%%%%%%%%%%%%%%%%%%%%%%%%%%%%%%%%%%%%%

%\cite{Hannestad:2010yi}
\bibitem{Hannestad:2010yi} 
  S.~Hannestad, A.~Mirizzi, G.~G.~Raffelt and Y.~Y.~Y.~Wong,
  %``Neutrino and axion hot dark matter bounds after WMAP-7,''  
JCAP {\bf 1008}, 001 (2010).
  %[arXiv:1004.0695 [astro-ph.CO]]. 
  %%CITATION = ARXIV:1004.0695;%%


\bibitem{Saito:2010pw} 
  S.~Saito, M.~Takada and A.~Taruya,
  %``Neutrino mass constraint with the Sloan Digital Sky Survey power spectrum of luminous red galaxies and perturbation theory,''  
  Phys.\ Rev.\ D {\bf 83}, 043529 (2011).
  %  [arXiv:1006.4845 [astro-ph.CO]].  
  %%CITATION = ARXIV:1006.4845;%%


%%%%%%%%%%%%total mass and Neff%%%%%%%%%%%%%%%%%%%%%%%%%%%%%%%%

%\cite{Crotty:2004gm}
\bibitem{Crotty:2004gm} 
  P.~Crotty, J.~Lesgourgues and S.~Pastor,
  %``Current cosmological bounds on neutrino masses and relativistic relics,''  
Phys.\ Rev.\ D {\bf 69}, 123007 (2004)  [hep-ph/0402049].  %%CITATION = HEP-PH/0402049;%%

%%%%%%%%%%%% + Ly alpha %%%%%%%%%%%%%%%
%\cite{Seljak:2006bg}
\bibitem{Seljak:2006bg} 
  U.~Seljak, A.~Slosar and P.~McDonald,
  %``Cosmological parameters from combining the Lyman-alpha forest with CMB, galaxy clustering and SN constraints,''  
JCAP {\bf 0610}, 014 (2006)  [astro-ph/0604335].  %%CITATION = ASTRO-PH/0604335;%%

\bibitem{Fukugita:2006rm} 
  M.~Fukugita, K.~Ichikawa, M.~Kawasaki and O.~Lahav,
  %``Limit on the Neutrino Mass from the WMAP Three Year Data,''  
  Phys.\ Rev.\ D {\bf 74}, 027302 (2006).
  %  [astro-ph/0605362].  
  %%CITATION = ASTRO-PH/0605362;%%

%%%%%%%%%%%%%%%%%%%%%%%%%%%%%%%%%%%%%%%%

%\cite{Komatsu:2008hk}
\bibitem{Komatsu:2008hk} 
  E.~Komatsu {\it et al.}  [WMAP Collaboration],
  %``Five-Year Wilkinson Microwave Anisotropy Probe (WMAP) Observations: Cosmological Interpretation,''  
Astrophys.\ J.\ Suppl.\  {\bf 180}, 330 (2009)  [arXiv:0803.0547 [astro-ph]].  %%CITATION = ARXIV:0803.0547;%%

%\cite{Reid:2009xm}
\bibitem{Reid:2009xm} 
  B.~A.~Reid, W.~J.~Percival, D.~J.~Eisenstein, L.~Verde, D.~N.~Spergel, R.~A.~Skibba, N.~A.~Bahcall and T.~Budavari {\it et al.},
  %``Cosmological Constraints from the Clustering of the Sloan Digital Sky Survey DR7 Luminous Red Galaxies,''  
Mon.\ Not.\ Roy.\ Astron.\ Soc.\  {\bf 404}, 60 (2010)  [arXiv:0907.1659 [astro-ph.CO]].  %%CITATION = ARXIV:0907.1659;%%

%\cite{Reid:2009nq}
\bibitem{Reid:2009nq} 
  B.~A.~Reid, L.~Verde, R.~Jimenez and O.~Mena,
  %``Robust Neutrino Constraints by Combining Low Redshift Observations with the CMB,''  
JCAP {\bf 1001}, 003 (2010)  [arXiv:0910.0008 [astro-ph.CO]].  %%CITATION = ARXIV:0910.0008;%%

%\cite{Hamann:2010pw}
\bibitem{Hamann:2010pw} 
  J.~Hamann, S.~Hannestad, J.~Lesgourgues, C.~Rampf and Y.~Y.~Y.~Wong,
  %``Cosmological parameters from large scale structure - geometric versus shape information,''
  JCAP {\bf 1007}, 022 (2010)
  [arXiv:1003.3999 [astro-ph.CO]].
  %%CITATION = ARXIV:1003.3999;%%
  

%\cite{Komatsu:2010fb}
\bibitem{Komatsu:2010fb} 
  E.~Komatsu {\it et al.}  [WMAP Collaboration],
  %``Seven-Year Wilkinson Microwave Anisotropy Probe (WMAP) Observations: Cosmological Interpretation,''  
Astrophys.\ J.\ Suppl.\  {\bf 192}, 18 (2011)  [arXiv:1001.4538 [astro-ph.CO]].  %%CITATION = ARXIV:1001.4538;%%


%%%%%%%%%%%%%%%%%%%%%only Neff%%%%%%%%%%%%%%%%%%%%%%

%\cite{Pierpaoli:2003kw}
\bibitem{Pierpaoli:2003kw} 
  E.~Pierpaoli,
  %``Constraints on the cosmic neutrino background,''  
Mon.\ Not.\ Roy.\ Astron.\ Soc.\  {\bf 342}, L63 (2003)  [astro-ph/0302465].  %%CITATION = ASTRO-PH/0302465;%%

%\cite{Crotty:2003th}
\bibitem{Crotty:2003th} 
  P.~Crotty, J.~Lesgourgues and S.~Pastor,
  %``Measuring the cosmological background of relativistic particles with WMAP,''  
Phys.\ Rev.\ D {\bf 67}, 123005 (2003)  [astro-ph/0302337].  %%CITATION = ASTRO-PH/0302337;%%



%\cite{Lesgourgues:2005yv}
\bibitem{Lesgourgues:2005yv} 
  J.~Lesgourgues, L.~Perotto, S.~Pastor and M.~Piat,
  %``Probing neutrino masses with cmb lensing extraction,''
  Phys.\ Rev.\ D {\bf 73}, 045021 (2006)
  [astro-ph/0511735].
  %%CITATION = ASTRO-PH/0511735;%%

\bibitem{dePutter:2009kn} 
  R.~de Putter, O.~Zahn and E.~V.~Linder,
  %``CMB Lensing Constraints on Neutrinos and Dark Energy,''  
Phys.\ Rev.\ D {\bf 79}, 065033 (2009).
%  [arXiv:0901.0916 [astro-ph.CO]].  
%%CITATION = ARXIV:0901.0916;%%


%%%%%%%%%%%%%%%%%%%%%%%%%%%%%%%%%%%%%%%%%%%%%%%%%%%%%%%%%%%%%%%%%%%%%%%%%%%%%%%%%%%%%%%%

\bibitem{Furlanetto:2006jb} 
  S.~Furlanetto, S.~P.~Oh and F.~Briggs,
  %``Cosmology at Low Frequencies: The 21 cm Transition and the High-Redshift Universe,''  
Phys.\ Rept.\  {\bf 433}, 181 (2006).
%  [astro-ph/0608032].  
%%CITATION = ASTRO-PH/0608032;%%


\bibitem{Loeb:2008hg} 
  A.~Loeb and S.~Wyithe,
  %``Precise Measurement of the Cosmological Power Spectrum With a Dedicated 21cm Survey After Reionization,''  
Phys.\ Rev.\ Lett.\  {\bf 100}, 161301 (2008).
%  [arXiv:0801.1677 [astro-ph]].  
%%CITATION = ARXIV:0801.1677;%%


\bibitem{Pritchard:2008wy} 
  J.~R.~Pritchard and E.~Pierpaoli,
  %``Constraining massive neutrinos using cosmological 21 cm observations,''  
Phys.\ Rev.\ D {\bf 78}, 065009 (2008).
%  [arXiv:0805.1920 [astro-ph]].  
%%CITATION = ARXIV:0805.1920;%%

\bibitem{Abazajian:2011dt} 
  K.~N.~Abazajian, E.~Calabrese, A.~Cooray, F.~De Bernardis, S.~Dodelson, A.~Friedland, G.~M.~Fuller and S.~Hannestad {\it et al.},
  %``Cosmological and Astrophysical Neutrino Mass Measurements,''  
Astropart.\ Phys.\  {\bf 35}, 177 (2011)  [arXiv:1103.5083 [astro-ph.CO]].  %%CITATION = ARXIV:1103.5083;%%


%%%%%%%%%%%%%%%%%%21cm line%%%%%%%%%%%%%%%%%%%%%%%%%%%%%%%%%%%%%%%%%%%%%%%%%%%%


%\cite{Pritchard:2011xb}
\bibitem{Pritchard:2011xb}
  J.~R.~Pritchard and A.~Loeb,
  %``21-cm cosmology,''
  arXiv:1109.6012 [astro-ph.CO].
  %%CITATION = ARXIV:1109.6012;%%  

%\cite{Madau:1996cs}
\bibitem{Madau:1996cs}
  P.~Madau, A.~Meiksin and M.~J.~Rees,
  %``21-cm Tomography of the Intergalactic Medium at High Redshift,''
  Astrophys.\ J.\  {\bf 475}, 429 (1997)
  [arXiv:astro-ph/9608010].
  %%CITATION = ASJOA,475,429;%%

%\cite{Furlanetto:2006tf}
\bibitem{Furlanetto:2006tf}
  S.~Furlanetto,
  %``The Global 21 Centimeter Background from High Redshifts,''
  Mon.\ Not.\ Roy.\ Astron.\ Soc.\  {\bf 371}, 867 (2006)
  [arXiv:astro-ph/0604040].
  %%CITATION = MNRAA,371,867;%%
  
%\cite{Pritchard:2008da}
\bibitem{Pritchard:2008da}
  J.~R.~Pritchard and A.~Loeb,
  %``Evolution of the 21 cm signal throughout cosmic history,''
  Phys.\ Rev.\  D {\bf 78}, 103511 (2008)
  [arXiv:0802.2102 [astro-ph]].
  %%CITATION = PHRVA,D78,103511;%%


%\cite{Wouthuysen:1952}
\bibitem{Wouthuysen:1952}
  S.~A.~Wouthuysen,
  %``,''
  Astron.\ J.\  {\bf 57}, 31 (1952).
  %%CITATION = ;%%


%\cite{Field:1958}
\bibitem{Field:1958}
  G.~B.~Field,
  %``,''
  Proc.\ IRE.\  {\bf 46}, 240 (1958).
  %%CITATION = ;%%


%\cite{Pritchard:2009zz}
\bibitem{Pritchard:2009zz}
  J.~R.~Pritchard and E.~Pierpaoli,
  %``Neutrino Mass From Cosmological 21 Cm Observations,''
  Nucl.\ Phys.\ Proc.\ Suppl.\  {\bf 188}, 31 (2009).
  %%CITATION = NUPHZ,188,31;%%
  

%\cite{Mao:2008ug}
\bibitem{Mao:2008ug}
  Y.~Mao, M.~Tegmark, M.~McQuinn, M.~Zaldarriaga and O.~Zahn,
  %``How accurately can 21 cm tomography constrain cosmology?,''
  Phys.\ Rev.\  D {\bf 78}, 023529 (2008)
  [arXiv:0802.1710 [astro-ph]].
  %%CITATION = PHRVA,D78,023529;%%

\bibitem{Joudaki:2011sv}
  S.~Joudaki, O.~Dore, L.~Ferramacho, M.~Kaplinghat and M.~G.~Santos,
  %``Primordial non-Gaussianity from the 21 cm Power Spectrum during the Epoch
  %of Reionization,''
  Phys.\ Rev.\ Lett.\  {\bf 107}, 131304 (2011)
  [arXiv:1105.1773 [astro-ph.CO]].
  %%CITATION = PRLTA,107,131304;%%

\bibitem{Chapman:2012yj}
 E.~Chapman, F.~B.~Abdalla, G.~Harker, V.~Jelic, P.~Labropoulos,
S.~Zaroubi, M.~A.~Brentjens and A.~G.~de Bruyn {\it et al.},
 %``Foreground Removal using FastICA: A Showcase of LOFAR-EoR,''
arXiv:1201.2190 [astro-ph.CO].  
%%CITATION = ARXIV:1201.2190;%%

\bibitem{LOFAR}
http://www.lofar.org/

%\cite{Lesgourgues:2006nd}
\bibitem{Lesgourgues:2006nd}
  J.~Lesgourgues and S.~Pastor,
  %``Massive neutrinos and cosmology,''
  Phys.\ Rept.\  {\bf 429}, 307 (2006)
  [arXiv:astro-ph/0603494].
  %%CITATION = PRPLC,429,307;%%


%\cite{Lesgourgues:2004ps}
\bibitem{Lesgourgues:2004ps} 
  J.~Lesgourgues, S.~Pastor and L.~Perotto,
  %``Probing neutrino masses with future galaxy redshift surveys,''
  Phys.\ Rev.\ D {\bf 70}, 045016 (2004)
  [hep-ph/0403296].
  %%CITATION = HEP-PH/0403296;%%

%\cite{MWA}
\bibitem{MWA}
  http://www.mwatelescope.org/
  %``
  %,''
  %%CITATION =;%%


%\cite{SKA}
\bibitem{SKA}
  http://www.skatelescope.org/
  %``
  %,''
  %%CITATION =;%%


%\cite{Tegmark:2009kv}
\bibitem{Tegmark:2009kv}
  M.~Tegmark and M.~Zaldarriaga,
  %``Omniscopes: Large Area Telescope Arrays with only N log N Computational
  %Cost,''
  Phys.\ Rev.\  D {\bf 82}, 103501 (2010)
  [arXiv:0909.0001 [astro-ph.CO]].
  %%CITATION = PHRVA,D82,103501;%%

%\cite{Wyithe:2007if}
\bibitem{Wyithe:2007if}
  S.~Wyithe and M.~F.~Morales,
  %``Biased Reionisation and Non-Gaussianity in Redshifted 21cm Intensity Maps
  %of the Reionisation Epoch,''
  arXiv:astro-ph/0703070.
  %%CITATION = ASTRO-PH/0703070;%%
  
%\cite{Bowman:2005cr}
\bibitem{Bowman:2005cr}
  J.~D.~Bowman, M.~F.~Morales and J.~N.~Hewitt,
  %``The Sensitivity of First Generation Epoch of Reionization Observatories and
  %Their Potential for Differentiating Theoretical Power Spectra,''
  Astrophys.\ J.\  {\bf 638}, 20 (2006)
  [arXiv:astro-ph/0507357].
  %%CITATION = ASJOA,638,20;%%  

%\cite{Tegmark:1996bz}
\bibitem{Tegmark:1996bz}
  M.~Tegmark, A.~Taylor and A.~Heavens,
  %``Karhunen-Loeve eigenvalue problems in cosmology: how should we tackle large
  %data sets?,''
  Astrophys.\ J.\  {\bf 480}, 22 (1997)
  [arXiv:astro-ph/9603021].
  %%CITATION = ASJOA,480,22;%%


%\cite{McQuinn:2005hk}
\bibitem{McQuinn:2005hk}
  M.~McQuinn, O.~Zahn, M.~Zaldarriaga, L.~Hernquist and S.~R.~Furlanetto,
  %``Cosmological Parameter Estimation Using 21 cm Radiation from the Epoch of
  %Reionization,''
  Astrophys.\ J.\  {\bf 653}, 815 (2006)
  [arXiv:astro-ph/0512263].
  %%CITATION = ASJOA,653,815;%%  
%\cite{Joudaki:2011sv}

%\cite{Takada:2005si}
\bibitem{Takada:2005si} 
  M.~Takada, E.~Komatsu and T.~Futamase,
  %``Cosmology with high-redshift galaxy survey: neutrino mass and inflation,''  
Phys.\ Rev.\ D {\bf 73}, 083520 (2006)  [astro-ph/0512374].  %%CITATION = ASTRO-PH/0512374;%%

%\cite{Slosar:2006xb}
\bibitem{Slosar:2006xb} 
  A.~Slosar,
  %``Detecting neutrino mass difference with cosmology,''  
Phys.\ Rev.\ D {\bf 73}, 123501 (2006)  [astro-ph/0602133].  %%CITATION = ASTRO-PH/0602133;%%

%\cite{DeBernardis:2009di}
\bibitem{DeBernardis:2009di} 
  F.~De Bernardis, T.~D.~Kitching, A.~Heavens and A.~Melchiorri,
  %``Determining the Neutrino Mass Hierarchy with Cosmology,''  
Phys.\ Rev.\ D {\bf 80}, 123509 (2009)  [arXiv:0907.1917 [astro-ph.CO]].  %%CITATION = ARXIV:0907.1917;%%


%\cite{Jimenez:2010ev}
\bibitem{Jimenez:2010ev}
  R.~Jimenez, T.~Kitching, C.~Pena-Garay and L.~Verde,
  %``Can we measure the neutrino mass hierarchy in the sky?,''
  JCAP {\bf 1005}, 035 (2010)
  [arXiv:1003.5918 [astro-ph.CO]].
  %%CITATION = JCAPA,1005,035;%%  


%\cite{Wong:2011ip}
\bibitem{Wong:2011ip} 
  Y.~Y.~Y.~Wong,
  %``Neutrino mass in cosmology: status and prospects,''
  Ann.\ Rev.\ Nucl.\ Part.\ Sci.\  {\bf 61}, 69 (2011)
  [arXiv:1111.1436 [astro-ph.CO]].
  %%CITATION = ARXIV:1111.1436;%%


%\cite{Chacko:2003dt}
\bibitem{Chacko:2003dt} 
  Z.~Chacko, L.~J.~Hall, T.~Okui and S.~J.~Oliver,
  %``CMB signals of neutrino mass generation,''
  Phys.\ Rev.\ D {\bf 70}, 085008 (2004)
  [hep-ph/0312267].
  %%CITATION = HEP-PH/0312267;%%



\bibitem{paper:POLARBEAR} 
H. Nishino et al, Proceedings of the 47th Rencontres de Moriond Cosmology (2012)


%\cite{Baumann:2008aq}
\bibitem{Baumann:2008aq} 
  D.~Baumann {\it et al.}  [CMBPol Study Team Collaboration],
  %``CMBPol Mission Concept Study: Probing Inflation with CMB Polarization,''
  AIP Conf.\ Proc.\  {\bf 1141}, 10 (2009)
  [arXiv:0811.3919 [astro-ph]].
  %%CITATION = ARXIV:0811.3919;%%



%References of CMB%%%%%%%%%%%%%%%%%%%%%%%%%%%

\bibitem{Lewis:1999bs} 
  A.~Lewis, A.~Challinor and A.~Lasenby,
  %``Efficient computation of CMB anisotropies in closed FRW models,''
  Astrophys.\ J.\  {\bf 538}, 473 (2000)  [astro-ph/9911177].
  %%CITATION = ASTRO-PH/9911177;%%

%\cite{CAMB}
\bibitem{CAMB}
  http://camb.info/
  %``
  %,''
  %%CITATION =;%%

%\cite{Okamoto:2003zw}
\bibitem{Okamoto:2003zw} 
  T.~Okamoto and W.~Hu,
  %``CMB lensing reconstruction on the full sky,''
  Phys.\ Rev.\ D {\bf 67}, 083002 (2003)[astro-ph/0301031].
  %%CITATION = ASTRO-PH/0301031;%%

\bibitem{paper:FUTURCMB}http://lpsc.in2p3.fr/perotto/


%%%%%%%%%%%%%%%%%%%%%%%%%%%%%%%%%%%%
\end{thebibliography}
\end{document}